\documentclass[reprint,amssymb, amsmath, aps, superscriptaddress, showpacs, footinbib, longbibliography, prl]{revtex4-1}
\usepackage{amssymb}
\usepackage{amsmath}
\usepackage{color}
\usepackage{graphicx}
\usepackage{footmisc,booktabs}
\usepackage{gensymb}
\usepackage{sidecap}
\usepackage{microtype}
\makeatletter
\newcommand{\RomanNumeralCaps}[1]
    {\MakeUppercase{\romannumeral #1}}
\makeatother
\begin{document}

\title{Field-induced instability of the quantum-spin-liquid ground state in the $J_{\rm eff}=\frac{1}{2}$ triangular-lattice compound NaYbO$_2$}

\author{K.~M.~Ranjith}
\email{ranjith.kumar@cpfs.mpg.de}
\affiliation{Max Planck Institute for Chemical Physics of Solids, 01187 Dresden, Germany}
\author{D.~Dmytriieva}
\affiliation{Hochfeld-Magnetlabor Dresden (HLD-EMFL), Helmholtz-Zentrum Dresden-Rossendorf, 01328 Dresden, Germany}
\affiliation{Institut f\"{u}r Festk\"orper- und Materialphysik, TU Dresden, 01062 Dresden Germany}

\author{S.~Khim}
\affiliation{Max Planck Institute for Chemical Physics of Solids, 01187 Dresden, Germany}
\author{J. Sichelschmidt}
\affiliation{Max Planck Institute for Chemical Physics of Solids, 01187 Dresden, Germany}
\author{S.~Luther}
\affiliation{Hochfeld-Magnetlabor Dresden (HLD-EMFL), Helmholtz-Zentrum Dresden-Rossendorf, 01328 Dresden, Germany}
\affiliation{Institut f\"{u}r Festk\"orper- und Materialphysik, TU Dresden, 01062 Dresden Germany}
\author{D. Ehlers}
\affiliation{Experimental Physics V, Center for Electronic Correlations and Magnetism, Institute of Physics, University of Augsburg, 86135 Augsburg, Germany}
\affiliation{Max Planck Institute for Chemical Physics of Solids, 01187 Dresden, Germany}

\author{H.~Yasuoka}
\affiliation{Max Planck Institute for Chemical Physics of Solids, 01187 Dresden, Germany}
\author{J.~Wosnitza}
\affiliation{Hochfeld-Magnetlabor Dresden (HLD-EMFL), Helmholtz-Zentrum Dresden-Rossendorf, 01328 Dresden, Germany}
\affiliation{Institut f\"{u}r Festk\"orper- und Materialphysik, TU Dresden, 01062 Dresden Germany}
\author{A.~A.~Tsirlin}
\affiliation{Experimental Physics VI, Center for Electronic Correlations and Magnetism, Institute of Physics, University of Augsburg, 86135 Augsburg, Germany}
\author{ H.~K\"{u}hne}
\affiliation{Hochfeld-Magnetlabor Dresden (HLD-EMFL), Helmholtz-Zentrum Dresden-Rossendorf, 01328 Dresden, Germany}
\author{M. Baenitz}
\email{michael.baenitz@cpfs.mpg.de}
\affiliation{Max Planck Institute for Chemical Physics of Solids, 01187 Dresden, Germany}

\date{\today}

\begin{abstract}\noindent
Polycrystalline samples of NaYbO$_2$ are investigated by bulk magnetization and specific-heat measurements, as well as by nuclear magnetic resonance (NMR) and electron spin resonance (ESR) as local probes. No signatures of long-range magnetic order are found down to 0.3~K, evidencing a highly frustrated spin-liquid-like ground state in zero field. Above 2\,T, signatures of magnetic order are observed in thermodynamic measurements, suggesting the possibility of a field-induced quantum phase transition. The $^{23}$Na NMR relaxation rates reveal the absence of magnetic order and persistent fluctuations down to 0.3~K at very low fields and confirm the bulk magnetic order above 2~T. The $H$-$T$ phase diagram is obtained and discussed along with the existing theoretical concepts for layered spin-$\frac{1}{2}$ triangular-lattice antiferromagnets.
\end{abstract}

\maketitle

Two-dimensional (2D) triangular-lattice antiferromagnets (TLAF) with spin-$\frac{1}{2}$ ions are a highly active field of research in current condensed-matter physics. Here, strong magnetic frustration due to the lattice geometry enhances quantum fluctuations and leads to exotic ground states. One of the most interesting examples is the quantum-spin-liquid (QSL) state, which is characterized by highly entangled spins, that remain dynamic down to zero temperature without any symmetry breaking~\cite{savary2016,balents2010,zhou2017}. Typical examples include $\kappa$-(BEDT-TTF)$_2$Cu$_2$(CN)$_3$, Et$_n$Me$_{4-n}$Sb[Pd(DMIT)$_2$]$_2$, and Ba$_3$CuSb$_2$O$_9$~\cite{Shimizu2003,yamashita2010,itou2008,Zhou2011}.
The presence of strong spin-orbit coupling (SOC), a local entanglement of spin and orbital degrees of freedom, further enhances the frustration and quantum fluctuations~\cite{chen2009,jackeli2009,ishizuka2014,lu2017}. Therefore, triangular systems with SOC are prime candidates for emerging QSL ground states~\cite{Maksimov2018,zhu2018}. Recently, YbMgGaO$_4$~\cite{li2015,li2015a} and NaYbS$_2$~\cite{baenitz2018} have been proposed as rare-earth triangular-lattice QSL candidates, where the 14-fold spin and orbital degeneracy of the Yb$^{3+}$ ions is lifted to four Kramers doublets due to the strong SOC and associated crystal electric field (CEF) effects. Eventually, the low-temperature behavior of these materials can be understood as arising from the effective (pseudo)spin-$\frac{1}{2}$ ground state. The first excited doublet is well separated from the ground state by an energy gap of $\Delta\sim$ 420~K and $\sim$200~K for YbMgGaO$_4$ and NaYbS$_2$, respectively. The magnetic entropy also supports the formation of a robust pseudospin-$\frac12$ state at low temperatures~\cite{li2015,li2015a,baenitz2018}.

It is known that in applied magnetic fields, the nature of the ground state is significantly modified for a spin-$\frac{1}{2}$ TLAF~\cite{chubokov1991}. The so-called ``order by disorder" mechanism in TLAF can stabilize different phases, including an oblique version of the 120\degree\ state ($Y$ phase), the collinear ``up-down-up" ($uud$) phase, and a canted version of the $uud$ phase (canted 2:1 phase)~\cite{chubokov1991}. The stabilization of the $uud$ phase is manifested by a magnetization plateau at $\frac{1}{3}M_s$ over a finite field range. Experimentally, this has been reported only in a few TLAFs, including Cs$_2$CuBr$_4$~\cite{Ono2003,Alicea2009} and Ba$_3$$M$Sb$_2$O$_9$ ($M$ = Co, Ni)~\cite{Fortune2009,Susuki2013,shirata2011}. Moreover, even a small interlayer interaction, which is present in most of the real materials, can generate additional frustration and quantum phase transitions at strong magnetic fields~\cite{gekht1997,Yamamoto2015}. Even though the non-spin-$\frac12$ TLAF families such as V$X_2 (X$ = Cl, Br) and $A$CrO$_2$ ($A$ = Li, Cu, Ag) have been widely discussed in this context,  the larger interlayer couplings and/or very strong in-plane couplings make the field range inaccessible for observing such field-induced quantum phase transitions experimentally~\cite{chubokov1991,gekht1997}.

Recently, rare-earth delafossite-type materials $A^{1+}R^{3+}X_2$, where $A$ is a monovalent nonmagnetic ion, $R$ is a rare-earth ion, and $X$ is a divalent chalcogen, have been proposed as triangular QSL candidates~\cite{baenitz2018,liu2018}. Most of them crystallize in the trigonal structures with $R\bar 3m$ symmetry, where the magnetic $R^{3+}$ ions form perfect triangular layers composed of edge-shared $R$X$_6$ octahedra. These triangular layers are separated by $A$X$_6$ octahedra along the crystallographic $c$ direction. In contrast to the heavily explored rare-earth TLAF QSL candidate YbMgGaO$_4$~\cite{li2016,shen2016,paddison2017,li2017a,zhang2018}, the $A^{1+}$Yb$^{3+}X_2$ delafossites are free from the intersite disorder, which may obscure the intrinsic properties of TLAFs~\cite{Xu2016,li2017b,zhu2017,Luo2017,Ma2018,Parker2018,Kimchi2018}.

A possible QSL ground state in NaYbO$_2$ is conjectured based on the absence of magnetic ordering and spin freezing down to 50~mK~\cite{liu2018,Ding2019}. Here, we utilize a range of different macroscopic and local techniques and elaborate on the magnetic behavior of NaYbO$_2$, including the peculiar field-induced order that appears above 2\,T. We first demonstrate that at low temperatures the Yb$^{3+}$ ions form effective spin-$\frac{1}{2}$ Kramers doublets due to the combined effect of CEF and SOC. The ground-state doublet is separated from the first excited state by a large energy gap of $\Delta\simeq$ 320~K (27~meV) as estimated from Yb ESR. The absence of magnetic long-range order (LRO) down to 0.3~K at zero or small applied fields, combined with the antiferromagnetic Curie-Weiss temperature of $\theta_{\rm CW}\sim$ $-6$~K, suggests a QSL-like ground state. Under external magnetic fields, NaYbO$_2$ undergoes several magnetic phase transitions that may be consistent with theoretical results for TLAF with small interlayer exchange interactions. In particular, we argue that the unusual behavior of NaYbO$_2$ arises from the proximity to an avoided antiferromagnetically ordered state. LRO is melted in zero field due to the combination of quantum fluctuations and exchange frustration, whereas an applied magnetic field quenches the quantum fluctuations and some of the exchange interactions and reinforces the LRO. Therefore, NaYbO$_2$ may be seen as a critical QSL due to the competition of weak magnetic exchange interactions and quantum fluctuations, as well as a unique system for exploring ground-state properties of triangular-lattice antiferromagnets with a strong SOC.

Polycrystalline samples of NaYbO$_2$ were synthesized using a solid-state reaction method, and the sample quality was confirmed by powder x-ray diffraction (for more details, see the Supplemental Material~\footnote{see the Supplemental Material}).  Magnetization ($M$) data were collected using a SQUID magnetometer (Quantum Design, VSM and MPMS) and an ac/dc susceptometer (PPMS). Studies below 2~K were performed using a $^3$He cooling stage (down to 500~mK) in the MPMS. Magnetization measurements in pulsed magnetic fields up to 35~T and at 500~mK were performed at the Dresden High Magnetic Field Laboratory (HLD). Specific-heat ($C_p$) measurements were performed on a pellet by using a relaxation technique (PPMS; Quantum Design) down to 350~mK. Electron spin resonance (ESR) experiments were performed at the $X$-band frequency (9.4~GHz). $T$-dependent nuclear magnetic resonance (NMR) experiments were carried out by applying  a pulsed NMR technique on the $^{23}$Na nucleus (nuclear spin $I$ = $\frac{3}{2}$) using a commercially available NMR spectrometer (Tecmag).

\begin{figure}[ht]
\includegraphics[clip,width=\columnwidth]{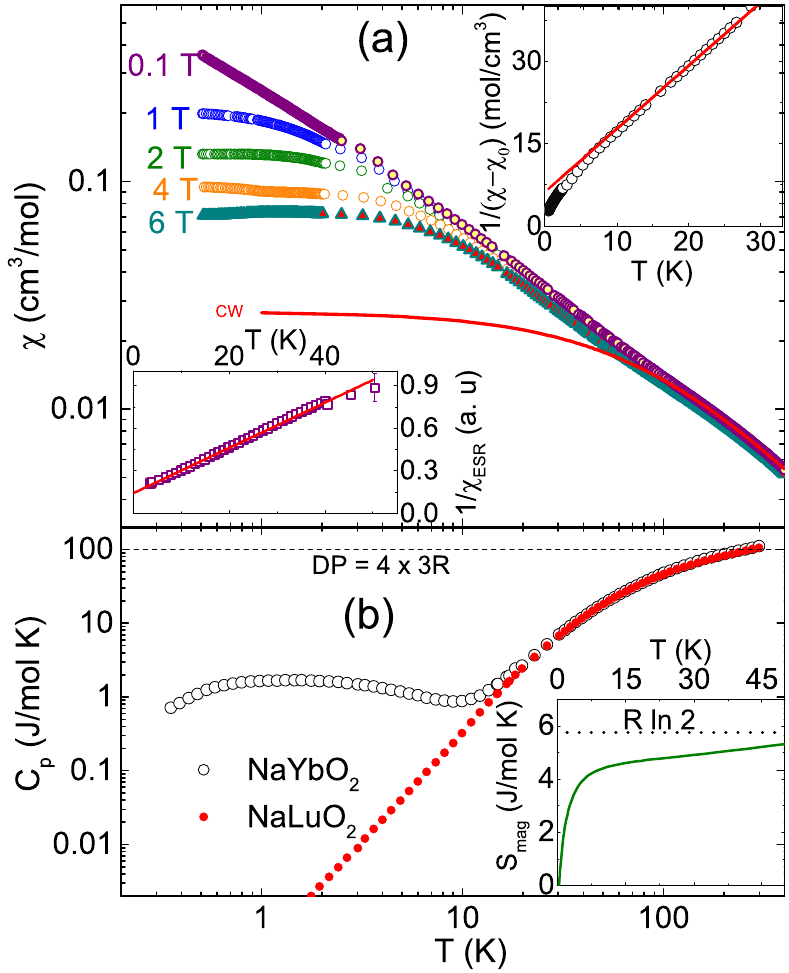}
\caption{(a)~$T$ dependence of the magnetic susceptibility $\chi(T)$ of NaYbO$_2$ measured at different applied magnetic fields. The solid line represents the high-temperature CW fit. The top inset shows the low-temperature data along with the CW fit of $\chi(T)$ at 0.1~T (after subtracting $\chi_0$). Bottom inset: Inverse ESR susceptibility 1/$\chi_{\rm ESR}$ below 50~K along with the CW fit. (b) Zero-field specific-heat data of NaYbO$_2$, in comparison with the specific-heat of the nonmagnetic structural analog NaLuO$_2$.  The horizontal dashed line indicates the Dulong-Petit value (DP). The inset shows the calculated magnetic entropy as a function of temperature and the dotted line indicates the expected high-temperature value of $R\ln 2$.}\label{Chi}
\end{figure}

Figure~\ref{Chi}(a)  shows the $T$ dependence of the magnetic susceptibility $\chi(T)$ of NaYbO$_{2}$, measured at different applied fields. Towards low temperatures, $\chi(T)$ increases in a Curie-Weiss (CW) manner. At low fields ($H\leq 2$~T), the $\chi(T)$ data do not show any indication of magnetic LRO down to 0.5~K. Above 200~K, 1/$\chi(T)$ can be well described using $\chi(T) =\chi_0+\frac{C}{T-\theta_{CW}}$, where $\chi_0$ is a $T$-independent contribution and the remaining term represents the CW law. Our fit yields $\chi_0\simeq$ 2$\times$10$^{-4}$~cm$^3$/mol, a Curie-Weiss temperature $\theta_{CW}=-100$~K, and an effective moment $\mu_{\rm eff} = 4.5$\,$\mu_{\rm B}$. The obtained parameters are consistent with the literature~\cite{liu2018,hashimoto2003}, and $\mu_{\rm eff}$ agrees with the theoretical prediction (4.54\,$\mu_{\rm B}$) for the Yb$^{3+}$ ion having a $^2F_{7/2}$ multiplet with $g$ = 8/7.

A change of slope was observed in 1/$\chi(T)$ below 100~K, indicating an evolution toward the ground-state CEF doublet with only a small intersite coupling. Below 25~K, 1/$\chi(T)$ could again be described by a Curie-Weiss law after subtracting a constant term $\chi_0 (\simeq$0.0065~cm$^3$/mol), yielding  $\mu_{\rm eff} = 2.6\mu_{\rm B}$ and $\theta_{CW}=-6$~K. Moreover, no indication of LRO is detected down to 0.35~mK, which yields a frustration parameter $f (\equiv\theta_{\rm CW}/T_{\rm N})\geq$ 17. The obtained moment $\mu_{\rm eff} (\equiv g\sqrt{J(J+1)}~\mu_{\rm B})$ is  reminiscent of an effective spin-$\frac{1}{2}$ state with an average $g$-value of 3. Similar effective spin-$\frac{1}{2}$ states are also reported for the Yb$^{3+}$-based TLAFs NaYbS$_{2}$~\cite{baenitz2018} and YbMgGaO$_{4}$~\cite{li2015}. At low temperatures, we observed a large $\chi_0$ value, which includes the Van-Vleck susceptibility $\chi_{\rm VV}$, arising from CEF excitations~\cite{Shirata2012}. Such a large low-temperature $\chi_{\rm VV}$ contribution is typical for Yb-based compounds~\cite{baenitz2018,miyasaka2009}.

Zero-field specific heat, measured as a function of temperature, is shown in Fig.~\ref{Chi}(b). No signatures of magnetic ordering were observed down to 0.35~K. The magnetic contribution to the specific heat is obtained by subtracting the lattice contribution, which is estimated by the specific heat of the isostructural reference compound NaLuO$_2$. The magnetic specific heat $C_{\rm mag}$ features a broad maximum at around 1.3~K with $C_{\rm mag}^{\rm max}/R\simeq 0.21$~\cite{Note1}. Below 1\,K, $C_{\rm mag}(T)$ follows a linear temperature dependence down to the lowest measured temperatures. Such a linear low-temperature dependence is rather unusual for 2D antiferromagnets, where magnon excitations would give a quadratic $T$-dependence~\cite{nakatsuji2005}. On the other hand, a linear low-$T$ dependence is common among various types of spin liquids~\cite{Zhou2011,yamashita2008,helton2007}. As shown in the inset of Fig.~\ref{Chi}(b), the magnetic entropy from lowest temperature to 40~K is nearly $~R\ln2$, as expected for an isolated Kramers doublet.

\begin{figure}[ht]
\includegraphics[clip,width=\columnwidth]{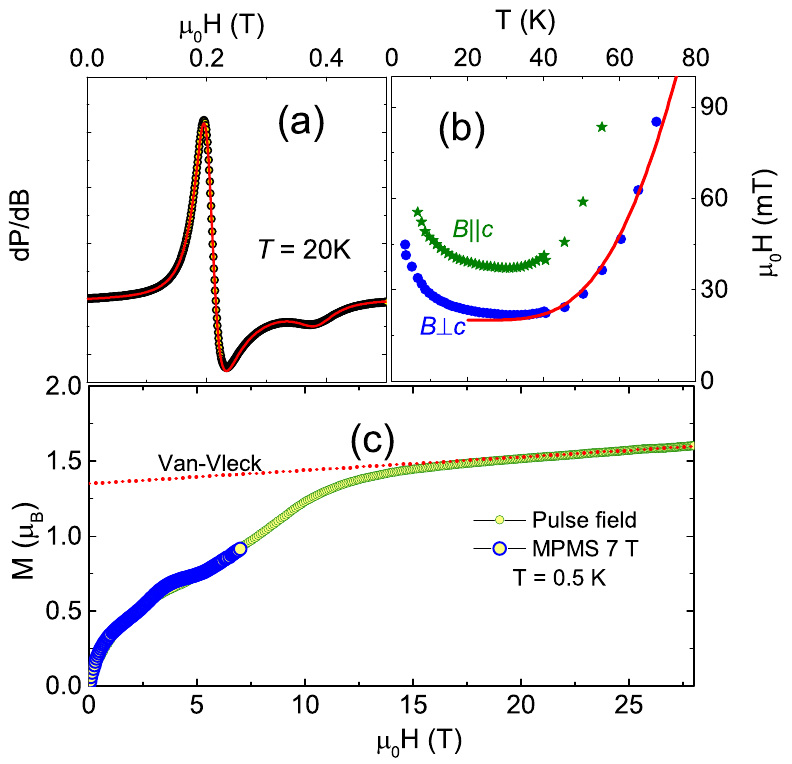}
\caption{(a) Derivative of the ESR absorption line of polycrystalline NaYbO$_{2}$ at 20 K. The solid line corresponds to a Lorentzian fit with powder-averaged uniaxial $g$-factor components. (b) $T$ dependence of the ESR linewidths $\Delta B$ with a fit (dashed line) toward high temperatures, indicating the influence of the first CEF level on the relaxation (see text). (c) Magnetization versus field for NaYbO$_2$, measured at $T$ = 0.5~K. The red dotted line indicates the Van Vleck contribution.}\label{MH}
\end{figure}

\begin{figure}[ht]
\includegraphics[clip,width=\columnwidth]{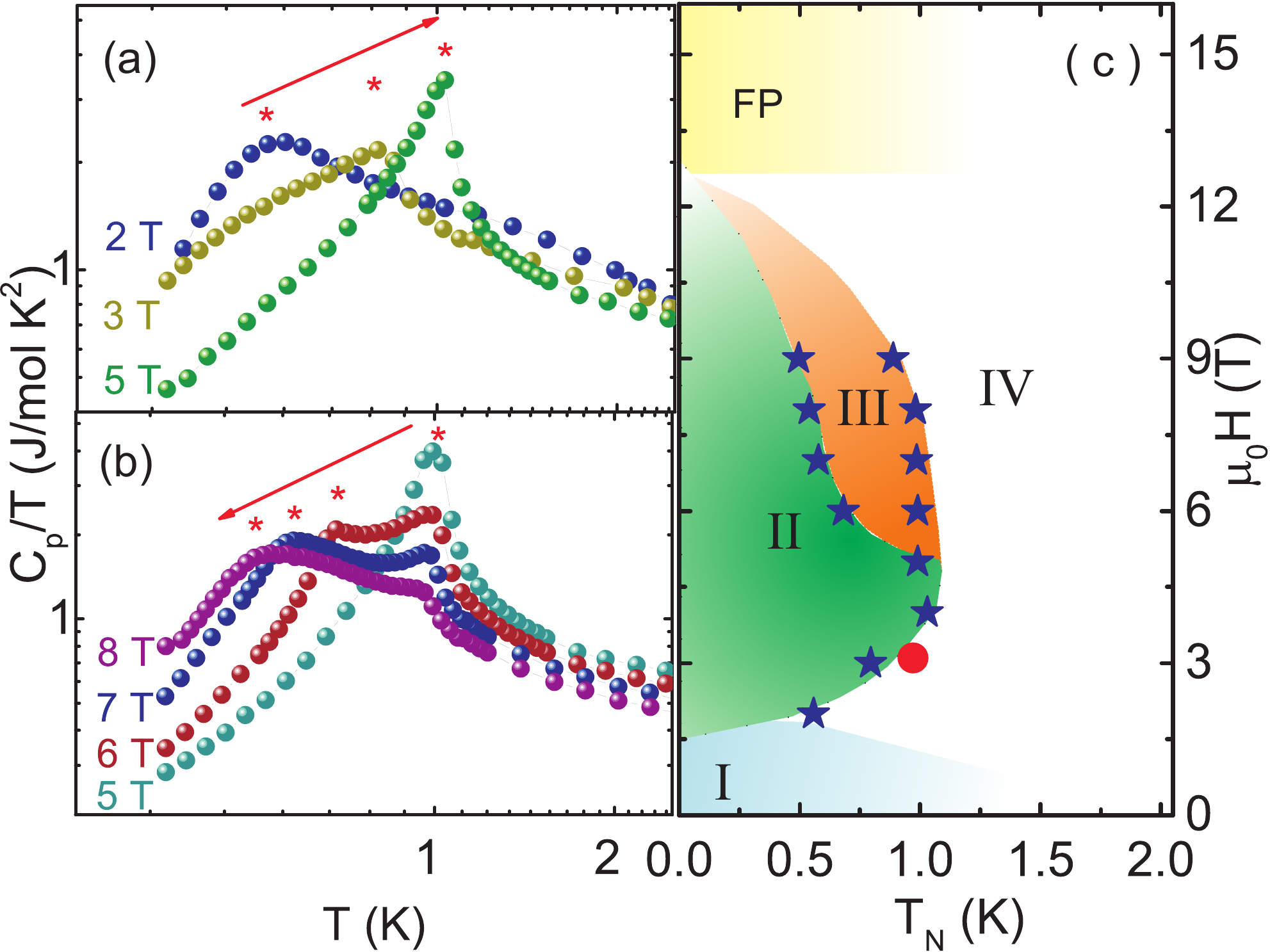}
\caption{(a),(b) $T$ dependence of the specific heat of NaYbO$_2$ measured at different applied fields. Red asterisk marks and arrows represent the magnetic transition and shift of the transition with applied field. (c) $H$-$T$ phase diagram with four different phases. The red circle corresponds to the NMR results at 3.1~T [35~MHz, Fig.\ref{nmr}(c)]. FP indicates the field-polarized phase.}\label{HT}
\end{figure}

ESR provides information about the local magnetic properties of Yb$^{3+}$ in NaYbO$_{2}$. The anisotropic $g$-factor of a polycrystalline sample was determined by fitting a Lorentzian ESR line
with powder averaging and a uniaxial $g$-factor anisotropy. As shown in Fig.~\ref{MH}(a), a narrow and well-resolved Yb-ESR line could be observed, very similar to the line observed for NaYbS$_{2}$ but quite different from the one reported for YbMgGaO$_{4}$~\cite{li2015a,baenitz2018}. In the latter compound, the averaged values of $g_\parallel$ = 3.06 and $g_\perp$ = 3.72 have been reported~\cite{li2015a}. However, the Mg-Ga site disorder triggers off-center displacements of the Yb$^{3+}$ ions and leads to a distribution of the $g$-values that manifests itself in the strong broadening of both the ESR line~\cite{li2015a} and CEF excitations as probed by inelastic neutron scattering~\cite{li2017b}.

In general, the ESR $g$-factor accounts for the Zeeman splitting of the lowest Kramers doublet of the Yb$^{3+}$ ion, and depends on the local site symmetry. The $g$-values for NaYbO$_2$ are found to be almost temperature-independent, but strongly anisotropic with $g_\parallel$ = 1.75(3) and $g_\perp$ = 3.28(8). The average $g$-value, $g_{\rm av} = \sqrt{(g_\parallel^2+2g_\perp^2)/3}\simeq$ 2.86, yields an effective moment of $\mu_{\rm ESR}\simeq$ 2.4~$\mu_{\rm B}$, which is in good agreement with the value obtained from the low-temperature CW fit~(2.6~$\mu_{\rm B}$). The integrated intensity of the ESR line is proportional to the local susceptibility. At low temperatures, $1/\chi_{\rm ESR}(T)$ yields $\theta_{CW}\simeq -9$~K, which is close to the low-temperature value extracted from the $1/\chi(T)$ analysis.

Figure~\ref{MH}(b) shows the ESR linewidth, $\Delta B(T)$, which for temperatures above 40~K is dominated by a broadening due to both spin-orbit coupling and modulation of the ligand field by lattice vibrations. The red-dashed line is the fit to the data with $\Delta B \propto 1/\exp(\Delta/T)-1$, which corresponds to an Orbach process, namely, the excited CEF states at an energy $\Delta$ above the ground state are involved in the relaxation \cite{orbach1961,abragam1970}. With a residual line width $\Delta B_{0}=15$~mT we obtain $\Delta\simeq320$~K ($\equiv 27$~meV).

The isothermal magnetization measured at $T=$ 0.5~K for magnetic fields up to 35~T, is shown in Fig.~\ref{MH}(c). It shows  two plateau-like anomalies at about $\sim$2 and $\sim$5~T before
reaching full saturation. This is rather unusual for a 2D TLAF, where one would expect a $\frac{M_s}{3}$ plateau only. In NaYbO$_2$, such multiple magnetization plateaus may indicate possible field-induced quantum phase transitions at low temperatures. $M(H)$ saturates nearly around 12~T and it linearly increases further in higher fields due to the Van-Vleck contribution. The estimated Van-Vleck contribution, $\chi_{\rm VV} \simeq 0.01~\mu_{\rm B}$/T $\simeq$ 0.0056~cm$^3$/mol is comparable with the value obtained in the low-$T$ $\chi(T)$ analysis.  The value of the saturation magnetization, $M_s\simeq$ 1.36~$\mu_{\rm B}$, is in good agreement with calculations based on the ESR $g$-value of $g_{\rm av}\simeq$ 2.86 and $J_{\rm eff}=\frac{1}{2}$, which leads to $M_s=gJ_{\rm eff}\mu_{\rm B}\sim$ 1.43~$\mu_{\rm B}$.

To gain more insight into the field-induced phases, we measured $\chi(T)$ and $C_{\rm p}(T)$ at different magnetic fields. The $\chi(T)$ data measured at higher fields ($H\geq2$~T) show a kink at low temperatures, which confirms the field-induced magnetic transitions (see Fig.~S3 in~\cite{Note1}). They are much better visible in the $C_{\rm p}(T)$ data shown in Fig.~\ref{HT}. The application of an external field of 2~T causes a magnetic transition at around 0.6~K, whereas no such transition was observed below 2~T. As shown in Fig.~\ref{HT}(a), this field-induced magnetic transition shifts towards higher temperatures with increasing magnetic field up to 4~T, where the transition temperature reaches its maximum at $T_{\rm N1}\simeq$ 1.03~K. Further increase in the field results in a decrease of $T_{\rm N1}$ and triggers an additional feature, related to a second transition at $T_{\rm N2}$ above 6~T. This kind of multiple magnetic transition is not uncommon in TLAF with easy-axis anisotropy~\cite{ranjith2016,ranjith2017}. As shown in Fig.~\ref{HT}(b), both transition temperatures, $T_{\rm N1}$ and $T_{\rm N2}$, decrease with further increase of the field.

\begin{figure}[ht]
\includegraphics[clip,width=\columnwidth]{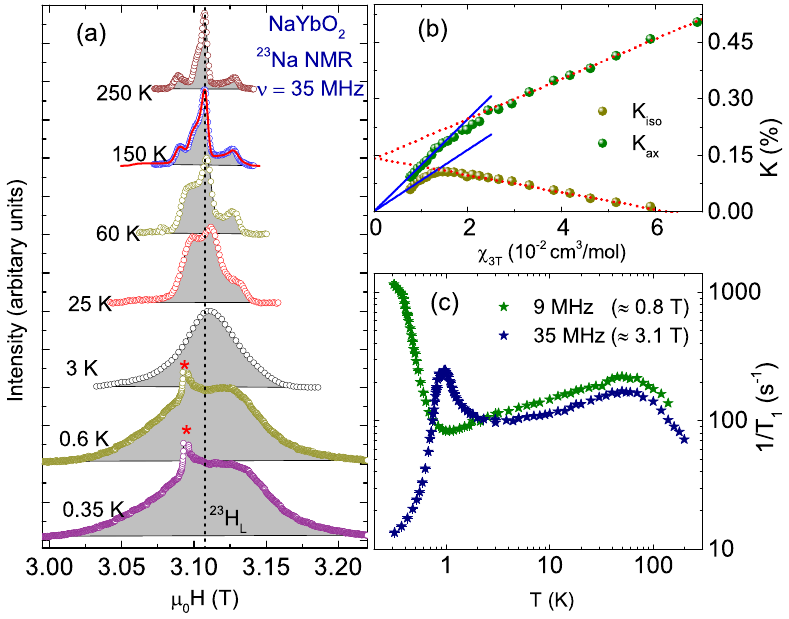}
\caption{(a)~$^{23}$Na NMR spectra measured at 35~MHz for different temperatures. (The vertical dotted line indicates the Larmor field, $^{23}H_{\rm L}$= 3.107~T, and asterisks mark the $^{63}$Cu signal from the coil used for the measurements at $^3$He temperatures.) (b) $K_{\rm iso}$ and $K_{\rm ax}$ vs $\chi$ at $\sim$3~T. The dotted and solid lines represent the fits at low and high temperatures, respectively. (c) $T$ dependence of 1/$T_1$ at two different fields. }\label{nmr}
\end{figure}

We further used $^{23}$Na NMR to probe the local static and dynamic magnetism of NaYbO$_2$. The $^{23}$Na NMR powder spectra, measured at 35~MHz, are shown in Fig.~\ref{nmr}(a). At high temperatures, the spectra exhibit a powder pattern typical for the $I=\frac{3}{2}$ nuclei with an anisotropic central line and two symmetric singularities, stemming from the first-order quadrupole interaction. The spectra can be well described by the isotropic and axially anisotropic NMR shift components $K_{\rm iso}$ and $K_{\rm ax}$, respectively, a quadrupole frequency $\nu_Q\simeq$ 456~kHz, and an electric-field gradient (EFG) asymmetry parameter $\eta=0$. The value of $\nu_Q$ is much larger than that observed in the isostructural NaYbS$_2$, which might be due to the smaller $c$ parameter of the oxide compound. The line is shifted and monotonously broadened with decreasing temperature, reflecting the anisotropic coupling of the nuclear moments to the electron moments of Yb$^{3+}$.

The $T$ dependence of the NMR shifts $K_{\rm iso}$ and $K_{\rm ax}$ could be extracted by consistently fitting these parameters to the spectra, taking broadening of the central and satellite lines into account [see the red solid line in Fig.~\ref{nmr}(a)]. The $T$ dependencies of the NMR shifts are remarkably different from those of $\chi(T)$~\cite{Note1}. In order to extract the transferred hyperfine coupling constant  $A_{\rm hf}$, $K_{\rm iso}$, and $K_{\rm ax}$ are plotted against the $\chi_{\rm 3T}$ as shown in Fig.~\ref{nmr}(b). The nonlinear dependence of $K$ vs $\chi$ indicates a $T$-dependent transferred hyperfine coupling in NaYbO$_2$.  $K_{\rm iso}$ and $K_{\rm ax}$ show linear dependencies with different slopes below about 60 and above 140~K.  The obtained high- and low-temperature hyperfine coupling constants are 456 Oe/$\mu_{\rm B}$ and $-127$~Oe/$\mu_{\rm B}$ for $K_{\rm iso}$, and 684~Oe/$\mu_{\rm B}$ and 296~Oe/$\mu_{\rm B}$ for $K_{\rm ax}$, respectively. The change in $A_{\rm hf}$ around 100~K might originate from the CEF splitting. At about 100~K, a depopulation of the excited CEF states may change the effective moment and hyperfine coupling. Similar anomalous $T$-dependent hyperfine coupling constants were also reported for other 4$f$-based spin systems~\cite{ohama1995,myers1973}. The NMR line is significantly broadened below 0.95~K, indicating the development of an internal field that confirms the onset of magnetic order, in agreement with our thermodynamic measurements.

The spin-lattice relaxation rate $1/T_1\propto[\sum_{\vec{q}}\mid A(\vec{q})\mid
^{2}\chi''(\vec{q},\omega_0)]$ probes low-energy spin excitations, where $A(\vec{q})$ is the form factor of the hyperfine interactions as a function of the wave vector $\vec{q}$, and $\chi''(\vec{q},\omega_0)$  is the imaginary part of the dynamic susceptibility at the nuclear Larmor frequency $\omega_0$.
Figure~\ref{nmr}(c) shows the $T$ dependence of the $^{23}$Na-spin-lattice relaxation rate $1/T_1$, measured at 0.8 and 3.1~T. At high temperatures, the overall temperature dependence of 1/$T_1$ is not affected by the field. The increase in $1/T_1$ upon cooling is followed by a broad maximum at around 70~K, instead of a $T$-independent behavior expected in the paramagnetic regime for a system with local magnetic moments. This anomalous behavior might be again attributed to the depopulation of the excited CEF levels. After passing through the maximum at 70~K, 1/$T_1$ decreases monotonously down to 3~K. At a small field of 0.8~T, 1/$T_1$ passes through a minimum at 1~K, followed by a strong increase toward low temperatures. This indicates a divergence of the imaginary part of the dynamic susceptibility ($\vec{q}\neq0$) on cooling and suggests some sort of critical fluctuations. In strong contrast, at 3.1~T the 1/$T_1$ shows a sharp anomaly at about $T_{\rm N1}\simeq$ 0.95~K, thus confirming the field-induced magnetic transition. Below $T_{\rm N}$, 1/$T_1$ decreases smoothly toward zero.

In the $1/T_1$ measurements, the magnetization recovery is fitted with a recovery function, which can provide information about the nature of the underlying relaxation mechanism~\cite{Note1}. The coefficient $\beta$ from a stretched exponential fit function is constant above 2~K, indicating a homogeneous distribution of fluctuating moments~\cite{Note1}. Below 2~K and at 3.1~T, $\beta$ changes and shows a dip at $T_{\rm N1}$, which is typical for strongly frustrated low-dimensional spin systems~\cite{itou2010,shockley2015,ranjith2018}. In contrast, at 0.8~T there is only a smooth change in $\beta$, which might indicate the onset of correlations among the Yb$^{3+}$ ions. At this point, it is unclear whether these are antiferromagnetic correlations in the vicinity of LRO, or if they are associated with a critical precursor state before entering the QSL.

A quantitative estimate of the in-plane exchange interaction $J$ can be made using the Heisenberg Hamiltonian $J\sum_{\langle i,j \rangle}{\mathbf S}_i{\mathbf S}_j$, where the summation is over bonds, and positive $J$ stands for an antiferromagnetic coupling. On the mean-field level, one expects the Curie-Weiss temperature of $\theta_{\rm CW} = -zJS(S+1)/3k_{\rm B}$, where $z$ is the number of nearest-neighbor spins and $k_{\rm B}$ is the Boltzmann constant. The number of nearest neighbors in a TLAF is $z=6$, so $\theta_{\rm CW}\simeq -6$~K results in an antiferromagnetic exchange interaction of $J/k_{\rm B}\simeq$ 4~K. The saturation for a purely 2D ($J'$ = 0) TLAF is
expected at $H_s = 4.5J\frac{k_B}{g\mu_B}\simeq$ 9.7~T, which is slightly below the observed value of 12~T. This indicates the presence of an interlayer interaction and/or of a second-neighbor interaction within the plane, both contributing to $H_s$.

Furthermore, the magnetic specific heat ($C_{\rm mag}$) of NaYbO$_2$ shows a rather broad maximum at around 2.1~K reaching $C_{\rm mag}^{\rm max}/R\simeq 0.21$. The magnetic specific heat is a direct measure of quantum fluctuations. For example, in 2D antiferromagnets $C_{\rm mag}^{\rm max}/R$ is reduced from 0.44 in the nonfrustrated square lattice~\cite{hofmann2003,johnston2000} to 0.22 in the geometrically frustrated triangular lattice~\cite{elstne1993,bernu2001}. Our result is thus very similar to the theoretical one for the triangular geometry. Moreover, the anticipated position of the maximum at $T/J=0.55$ (2.1\,K in NaYbO$_2$) perfectly fits our data. It is worth noting that $C_{\rm mag}$ of YbMgGaO$_4$ also shows a maximum around the same temperature, but its magnitude is higher, whereas the average nearest-neighbor coupling $\bar J\simeq 1.4$\,K is lower than in NaYbO$_2$. Therefore, NaYbO$_2$ seems to be much closer to the Heisenberg TLAF than YbMgGaO$_4$.

Despite this proximity, NaYbO$_2$ does not show the transition toward the $120^{\circ}$ ordered state expected in a Heisenberg TLAF. Instead, we detect signatures of field-induced magnetic order that have not been observed in other Yb$^{3+}$-based triangular-lattice magnets reported to date. In Fig.~\ref{HT}(c), we present a comprehensive $H$-$T$ phase diagram, where four different phases (\RomanNumeralCaps{1}-\RomanNumeralCaps{4}) may occur.  Phase \RomanNumeralCaps{4} is the paramagnetic phase, whereas phase \RomanNumeralCaps{1} represents a fluctuating QSL-like state, as reported by Liu $et~al.$~\cite{liu2018}. Phases \RomanNumeralCaps{2} and \RomanNumeralCaps{3} represent the magnetically ordered phases. This phase diagram suggests that NaYbO$_2$ may be at the verge of a field-induced quantum critical point, and indeed we observed signatures of critical fluctuations in $1/T_1$. Further on, the NMR spectra measured at $\sim$7~T show different shapes within phase \RomanNumeralCaps{2} and phase \RomanNumeralCaps{3} suggesting that  these are two different ordered states.
Neutron-diffraction work under external magnetic fields would be highly desirable to understand the exact nature of these ordered phases.

In conclusion, we have studied the rare-earth triangular-lattice QSL candidate NaYbO$_2$ in detail. The low-temperature behavior of the Yb$^{3+}$ ions can be attributed to a spin-$\frac{1}{2}$ Kramers doublet, well separated from the first excited doublet. The results of the $^{23}$Na NMR reveal increasing magnetic fluctuations toward low temperatures and suggest that NaYbO$_2$ may be a critical spin liquid in analogy to the critical Fermi liquid. The instability of this tentative QSL state is observed in higher fields, with two distinct ordered states forming above 2\,T. Upon applying magnetic fields, the fluctuations are suppressed, and the system undergoes a quantum phase transition. The observed anomalous temperature dependence of the NMR shifts and spin-lattice relaxation rates are ascribed to CEF effects. Furthermore, thermodynamic properties are in agreement with theoretical predictions for spin-$\frac{1}{2}$ triangular-lattice Heisenberg antiferromagnets. Therefore, NaYbO$_2$ could be a perfect model system for understanding the spin-$\frac{1}{2}$ magnetism with spin-orbit coupling in the triangular geometry~\cite{Maksimov2018,zhu2018}. In addition, it features critical fluctuations that deserve detailed investigations with local probes such as muon
spin rotation/relaxation or inelastic neutron scattering.

\emph{Note added}. Recently, we became aware of Ref.~\cite{Bordelon2019}, where NaYbO$_2$ was further investigated by neutron-diffraction experiments.

We thank H.~Rosner, B.~Schmidt, A. Mackenzie, C.~Geibel, L. Hozoi, and D. Inosov for fruitful discussions. We thank C.~Klausnitzer and Y. Skourski for technical support. We acknowledge the support from the Deutsche Forschungsgemeinschaft (DFG) through SFB 1143, and support by the HLD at HZDR, member of the European Magnetic Field Laboratory (EMFL).
A.A.T. acknowledges financial support by the Federal Ministry for Education and Research through the Sofja Kovalevskaya Award of Alexander von Humboldt Foundation.

\end{document}